\documentclass[reprint,superscriptaddress,nobibnotes,amsmath,amssymb,twocolumn,prl]{revtex4-1}

\usepackage{pdfpages}
\usepackage{etoolbox} 

\usepackage{graphicx}
\usepackage{bm} % bold math
\usepackage{mathptmx}
\usepackage{color}
\usepackage[caption=false]{subfig}
\usepackage{mathtools} % loads amsmath

\usepackage{hyperref}
\hypersetup{colorlinks=true, linkcolor=blue, citecolor=blue, urlcolor=blue}

\graphicspath{{./img/}}

\newcommand{\beq}[1]{\begin{equation}\label{#1}}
\newcommand{\eep}{\;.\end{equation}}
\newcommand{\eec}{\;,\end{equation}}
\newcommand{\eeq}{\end{equation}}

\newcommand*\dd{\mathop{}\!\mathrm{d}} %differential d
\newcommand{\bvec}[1]{\mathbf{#1}}

\newcommand{\lb}{\left(}
\newcommand{\rb}{\right)}

\newcommand*\chem[1]{\ensuremath{\mathrm{#1}}} % for chemical symbols

\renewcommand{\a}{\alpha}
\renewcommand{\b}{\beta}
\renewcommand{\k}{\kappa}
\renewcommand{\th}{\theta}

\newcommand{\D}{\Delta}

\DeclareMathAlphabet{\mathcal}{OMS}{cmsy}{m}{n} % Changes font for mathcal but leaves the rest of the math fonts in Times.

\newcommand{\Ef}{\mathcal{E}}   % Electric field
\newcommand{\Vstack}{\mathcal{V}_{\text{stack}}} 
\newcommand{\Ct}{\mathcal{C}_3}

\makeatletter
\patchcmd{\@outputpage@head}{\@ifx{\LS@rot\@undefined}{}{\LS@rot}}{}{}{}
\makeatother

\begin{document}

%%% TITLE, AUTHORS, ABSTRACT%%%

\title{Polar meron-antimeron networks in strained and twisted bilayers}

\author{Daniel Bennett}
\email{dbennett@seas.harvard.edu}
\affiliation{Physique Théorique des Matériaux, QMAT, CESAM, University of Liège, B-4000 Sart-Tilman, Belgium}
\affiliation{Theory of Condensed Matter Group, Cavendish Laboratory, University of Cambridge, J.\,J.\,Thomson Avenue, Cambridge CB3 0HE, United Kingdom}
\affiliation{John A.~Paulson School of Engineering and Applied Sciences, Harvard University, Cambridge, Massachusetts 02138, USA}

\author{Gaurav Chaudhary}
\affiliation{Theory of Condensed Matter Group, Cavendish Laboratory, University of Cambridge, J.\,J.\,Thomson Avenue, Cambridge CB3 0HE, United Kingdom}

\author{Robert-Jan Slager}
\affiliation{Theory of Condensed Matter Group, Cavendish Laboratory, University of Cambridge, J.\,J.\,Thomson Avenue, Cambridge CB3 0HE, United Kingdom}

\author{Eric Bousquet}
\affiliation{Physique Théorique des Matériaux, QMAT, CESAM, University of Liège, B-4000 Sart-Tilman, Belgium}

\author{Philippe Ghosez}
\affiliation{Physique Théorique des Matériaux, QMAT, CESAM, University of Liège, B-4000 Sart-Tilman, Belgium}

\begin{abstract}

\begin{center}
\textbf{Abstract}
\end{center}

Out-of-plane polar domain structures have recently been discovered in strained and twisted bilayers of inversion symmetry broken systems such as hexagonal boron nitride. Here we show that this symmetry breaking also gives rise to an in-plane component of polarization, and the form of the total polarization is determined purely from symmetry considerations. The in-plane component of the polarization makes the polar domains in strained and twisted bilayers topologically non-trivial, forming a network of merons and antimerons (half-skyrmions and half-antiskyrmions). For twisted systems, the merons are of Bloch type whereas for strained systems they are of Néel type. We propose that the polar domains in strained or twisted bilayers may serve as a platform for exploring topological physics in layered materials and discuss how control over topological phases and phase transitions may be achieved in such systems.
\end{abstract}

\maketitle

\section{Introduction}

\begin{figure*}[t!]
\centering
\includegraphics[width=\linewidth]{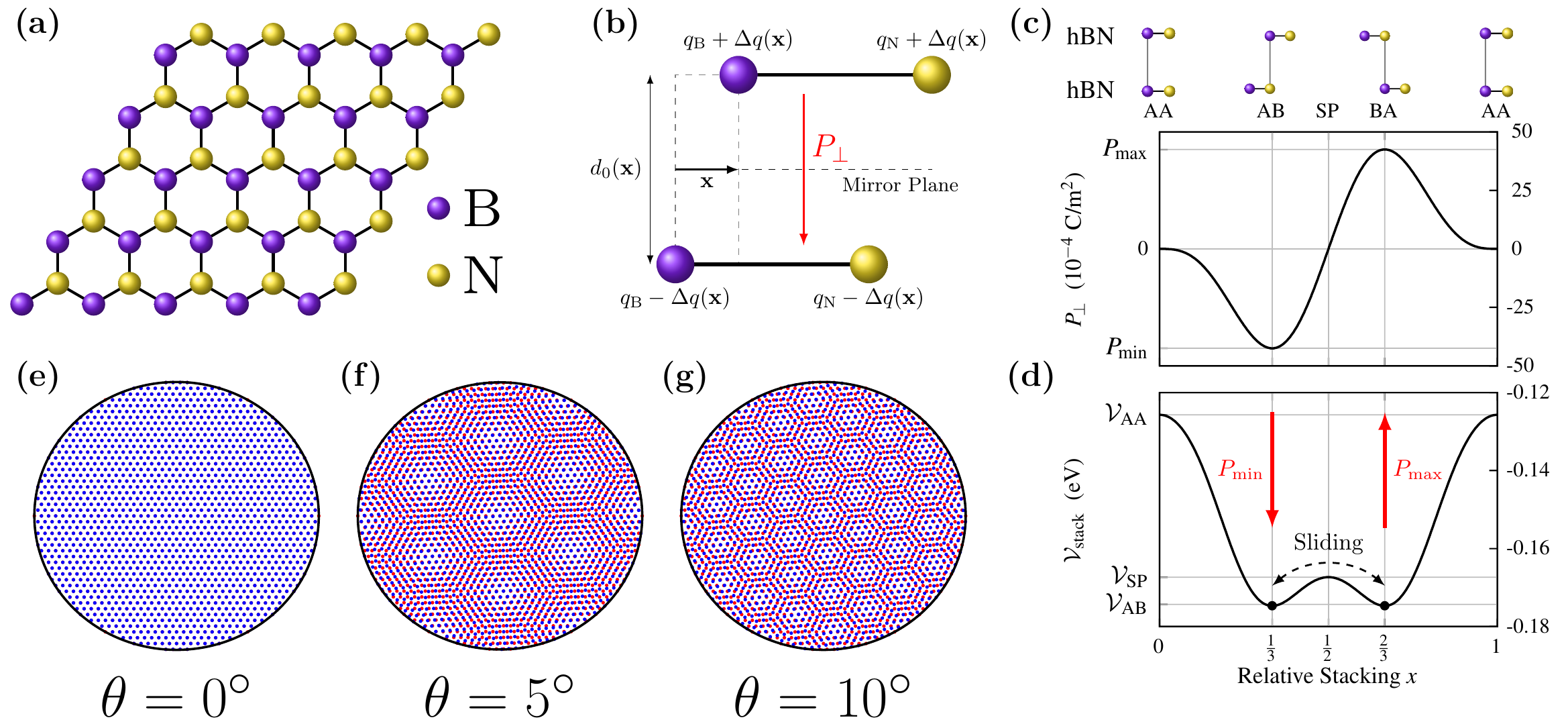}
\caption{\textbf{Ferroelectricity in twisted bilayers.} \textbf{(a)}: Sketch of a monolayer of hBN. \textbf{(b)}: Illustration of the interlayer charge transfer $\D q(\bvec{x})$ and resulting polarization arising from relative sliding $\bvec{x}$ in bilayer hBN, where $q_{\text{B}}$ and $q_{\text{N}}$ are the charges of the B and N atoms, respectively, and $d_0(\bvec{x})$ is the equilibrium layer separation. \textbf{(c)}: Out-of-plane polarization $P_{\perp}$ and \textbf{(d)} stacking energy $\Vstack$ in bilayer hBN as a function of relative stacking $x$ along the unit cell diagonal, from first-principles calculations in Ref.~\cite{bennett2022theory}. The stacking configurations at the extrema, AA, AB, SP and BA are labeled and sketched above the plots, and their stacking energies $\mathcal{V}_{\text{AA}}$, $\mathcal{V}_{\text{SP}}$ and $\mathcal{V}_{\text{AB}} = \mathcal{V}_{\text{BA}}$ are marked on the vertical axis of \textbf{(d)}. Changing between the energetically stable AB and BA stacking configurations via vdW sliding inverts the polarization between the minimum and maximum values $P_{\text{min}}$ and $P_{\text{max}}$. \textbf{(e)-(g)}: Sketch of red and blue hexagonal bilayers with relative twist angles of, $\theta=0^{\circ},5^{\circ},10^{\circ}$, respectively.}
\label{fig:pol-hbn-intro}
\end{figure*}

Recently it has been realized that ferroelectricity can occur in layered systems comprised of stacks of two-dimensional (2D) materials such as hexagonal boron nitride (hBN), see Fig.~\ref{fig:pol-hbn-intro} (a), provided the stack of layers does not have inversion symmetry \cite{li2017binary}. In an aligned stack of hBN (3R stacking), which has four non-orthogonal mirror planes and is therefore non-centrosymmetric but still non-polar, sliding one layer over the other breaks the mirror symmetry about the plane which is parallel to and half-way between the layers, resulting in an interlayer transfer of electronic charge and an out-of-plane polarization \cite{li2017binary,bennett2022electrically,bennett2022theory}, see Fig.~\ref{fig:pol-hbn-intro} (b). For anti-aligned hBN (2H stacking), there is an inversion center for every stacking, and the system is nonpolar. Applying an electric field to aligned hBN, the polarization can be inverted via a relative sliding between the layers (van der Waals sliding)  \cite{stern2020interfacial,yasuda2021stacking} in order to align the polarization with the field, see Fig.~\ref{fig:pol-hbn-intro} (c). This mechanism is highly unconventional when compared to the ferroelectricity observed in \chem{ABO_3} oxide perovskites, in particular because the polarization generated is perpendicular to the atomic motion.

In a twisted bilayer, two layers are twisted with respect to one another, forming a supercell known as a moir\'e superlattice, see Figs.~\ref{fig:pol-hbn-intro} (d)-(f). A moir\'e superlattice can also be generated by introducing a small relative strain or lattice mismatch between the layers. Twisting has been shown to result in novel phenomena such as superconducting \cite{cao2018unconventional} and insulating \cite{cao2018correlated} behavior in bilayer `magic angle graphene', and recently ferroelectricity in hBN \cite{zheng2020unconventional,yasuda2021stacking,woods2021charge}. A small lattice mismatch moir\'e superlattice formed between non-Bravais lattice monolayers has local regions with different stacking configurations, which may locally break mirror symmetry. This symmetry breaking in conjugation with absence of inversion center in the monolayer leads to local out-of-plane polarization with stacking dependent direction \cite{bennett2022electrically,bennett2022theory}. Thus, the stacking domains in strained and twisted bilayers can be identified as out-of-plane `moir\'e polar domains' (MPDs). The experimentally observed ferroelectricity has been attributed to the motion of the domain walls separating the MPDs in response to an applied out-of-plane electric field. As a result, the MPDs with polarization (anti-)aligned to the field (shrink) grow in size.

Something which to our knowledge has not been considered is the possibility of an \textit{in-plane} polarization, both in moir\'e superlattices and commensurate layered systems. Because layered systems are periodic in the in-plane directions, the in-plane polarization is a lattice valued quantity, and only changes in the in-plane polarization are well defined, modulo a quantum of polarization \cite{king1993theory}. 2D honeycomb compounds with an AB sublattice structure have a triangular in-plane polarization lattice \cite{bristowe2013one}, and it is natural to expect that changing the stacking configuration in a bilayer may result in a continuous change in the in-plane polarization.

We show with first-principles calculations of bilayer hBN that an in-plane polarization is indeed generated in layered system when one layer slides over the other. As a consequence, the MPDs do not just point in the out-of-plane direction, but also have an intricate in-plane component, such that the polarization vector has topologically non-trivial winding and the MPDs form a network of merons and antimerons (winding numbers  $\pm\frac{1}{2}$). This indicates that the polar properties of layered systems can have rich topological structure. Topological polar structures such as skyrmions \cite{das2019observation,han2022high} and merons \cite{wang2020polar} have been observed in ferroelectric materials such as oxide perovskites, and have been shown to result in novel physics of interest for future applications in nanotechnology, such as negative capacitance \cite{zubko2016negative} and high density information processing \cite{han2022high}. So far, band topology in the moir\'e systems has appeared in the electronic structure of magic angle graphene \cite{song2019all,Lu2021}, Chern bands in twisted topological insulators \cite{Lian2020} , and topological superconductivity in twisted cuprates \cite{can2021high}.
The polar meron-antimeron network suggests that moir\'e materials also exhibit real space  topology, echoing recent similar discoveries of topologically nontrivial strain fields in twisted bilayers \cite{engelke2022non} and magnetic textures in moir\'e patterned topological insulators \cite{guerci2022designer}.

\section{Results}

\begin{figure*}[t!]
\centering
\includegraphics[width=\linewidth]{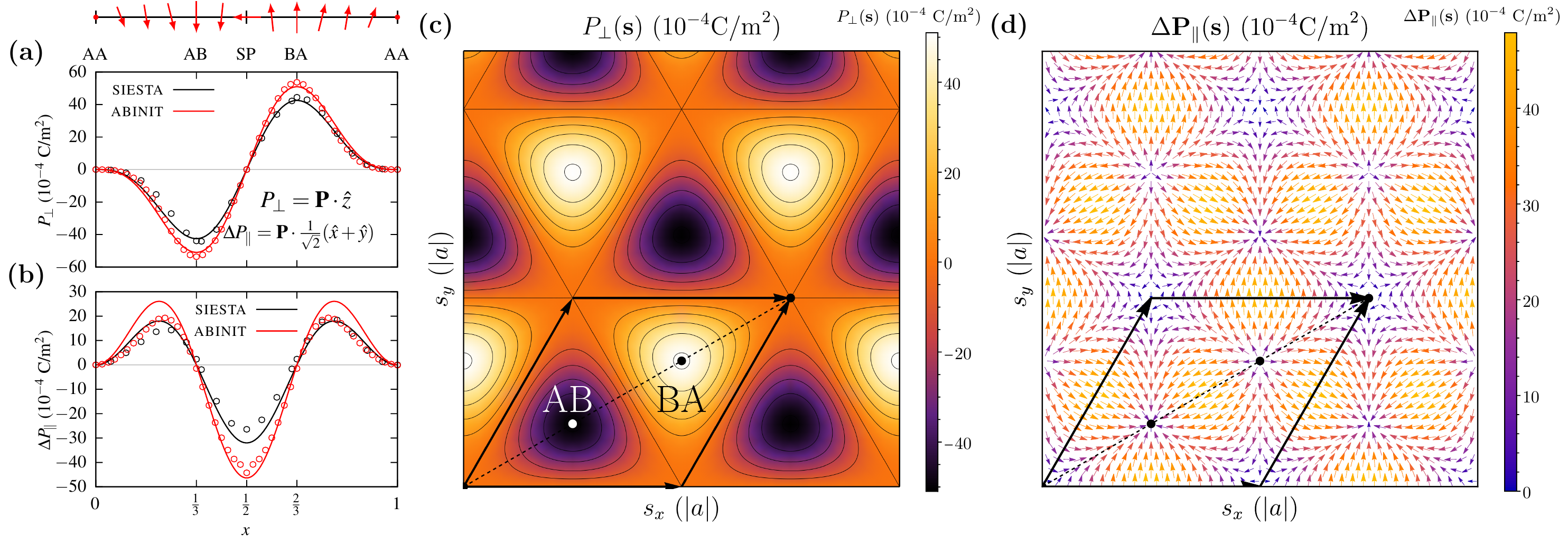}
\caption{\textbf{Total polarization in bilayer hBN.} \textbf{(a)} Out-of-plane polarization $P_{\perp}$ and \textbf{(b)} change in in-plane polarization $\D P_{\parallel}$ in fractional coordinates along the configuration space diagonal, calculated with {\sc siesta} (black) and {\sc abinit} (red), where $\hat{x}$ and $\hat{y}$ are the normalized lattice vectors of the bilayer, and $\hat{z}$ is the unit vector normal to the bilayer. The hollow points are first-principles measurements and the solid lines are fitting to $\Ct$ symmetric basis functions. The winding of the total polarization along the configuration space diagonal is sketched above. 2D plot of \textbf{(c)} out-of-plane and \textbf{(d)} in-plane polarization $\D \bvec{P}_{\parallel}$ in configuration space, in Cartesian coordinates $\bvec{s} = (s_x,s_y)$ and $|a|$ is the bilayer lattice constant. The black arrows represent the commensurate bilayer lattice vectors.}
\label{fig:pol-main}
\end{figure*}

We first calculate the out-of-plane and in-plane polarization of 3R-stacked bilayer hBN, using the first-principles {\sc siesta} \cite{siesta} and {\sc abinit} \cite{gonze2009abinit} codes (see Methods). The polarization is calculated in commensurate 3R-stacked bilayer hBN as a function of relative displacement $\bvec{s}$ between the layers, i.e.~in configuration space \cite{carr2018relaxation}, which can be used to estimate the polarization in real space for arbitrary strains and twist angles, provided the mismatch is small enough that the local stackings are well approximated by a commensurate bilayer plus a relative translation (see Supplementary Material, Section I). A changing in-plane polarization was found, of the same order of magnitude as the out-of-plane polarization, see Figs.~\ref{fig:pol-main} (a) and (b).

The shape of the polarization field as a function of relative stacking is determined purely from symmetry considerations, although the magnitude is material specific. The aligned AA stacking with space group P$\bar{6}$m2 ({\#}187) has three out-of-plane mirror planes running through $\Ct$ rotations of the ${\hat{x}+\hat{y}}$ unit cell diagonal, where $\hat{x} = \begin{bmatrix} 1 \\ 0 \end{bmatrix}$ and $\hat{y} = \begin{bmatrix} 1/2 \\ \sqrt{3}/2 \end{bmatrix}$, and an in-plane mirror plane half way through the layers. The bilayer is therefore non-polar for this stacking, but because the mirror planes are not orthogonal, it is not centrosymmetric. Sliding one layer over the other by $\frac{1}{3}$ or $\frac{2}{3}$ along the ${\hat{x}+\hat{y}}$ unit cell diagonal (or one of its $\Ct$ rotations), the energetically favourable AB and BA stackings are realized, both with the polar space group P3m1 ({\#}156). The three mirror symmetries through $\Ct$ rotations of the unit cell diagonal are preserved, but the in-plane mirror symmetry is broken, allowing for a polarization only in the out-of-plane direction. Half way between the AB and BA stacking configurations, at the saddle point (SP, $x=\frac{1}{2}$), the Abm2 ({\#}39) space group is realized, which only has an out-of-plane mirror symmetry through the ${\hat{x}+\hat{y}}$ unit cell diagonal, assuming the relative translation of the layers is along this diagonal. Additionally, the system is left invariant after mirroring about the plane half-way between the layers plus a non-symmorphic translation of $\frac{1}{2}\lb{\hat{x}+\hat{y}}\rb$, preventing any out-of-plane polarization. Thus, only an in-plane polarization along the ${\hat{x}+\hat{y}}$ unit cell diagonal is allowed. For any other translation along the ${\hat{x}+\hat{y}}$ unit cell diagonal or one of its $\Ct$ rotations, the bilayer has the Cm ({\#}8) space group, with only the mirror plane running through that diagonal. The polarization is then confined to that mirror plane, but can have both in-plane and out-of-plane components. Finally, for a translation not along the unit cell diagonal, the P1 ({\#}1) space group is realized, and the polarization can point in any direction.

Using the in-plane mirror symmetry of the AA stacking configuration, we can further deduce that the out-of-plane polarization $P_{\perp}$ must be an odd function of in-plane translations (see Supplementary Material, Section I). Requiring also that $P_{\perp}$ transforms as a scalar field with respect to $\Ct$ rotations about the out-of-plane axes through AA, AB and BA, it can be shown to be of the form
\beq{eq:fit-Pz}
P_{\perp}(x,y) = P^{\text{odd}}_1 \bigg[  \sin{(2\pi x)} + \sin{(2\pi y)} - \sin{(2\pi (x+y))} \bigg]
\eeq
in fractional coordinates $\bvec{x}$ in configuration space, where $(x,y) \equiv \bvec{x}$ are fractions of the lattice vectors $\hat{x}$ and $\hat{y}$; the fractional and Cartesian coordinates in configuration space are related via ${\bvec{s}=g\bvec{x}}$, where ${g = \begin{bmatrix} 1 & 1/2 \\ 0 & \sqrt{3}/2\end{bmatrix}}$. We can also show that the in-plane polarization $\bvec{P}_{\parallel}$, must be even with respect to in-plane translations. Additionally, it should transform like a vector with with respect to $\Ct$ rotations about the out-of-plane axes, and therefore must be of the form
\beq{eq:fit-Px}
\D \bvec{P}_{\parallel}(x,y) = P^{\text{even}}_1 
\begin{bmatrix}  
\cos{(2\pi x)} - \cos{(2\pi (x+y))}  \\
\cos{(2\pi y)} - \cos{(2\pi (x+y))}  \\
\end{bmatrix}
\eep
Eqs.~\eqref{eq:fit-Pz} and \eqref{eq:fit-Px} were fit to the 1D data along the ${\hat{x}+\hat{y}}$ unit cell diagonal in Figs.~\ref{fig:pol-main} (a) and (b). The 1D data were sufficient to obtain a good fit to the 2D functions, which was verified by fitting to a larger set of displacements parametrizing the entire unit cell in configuration space. The out-of-plane and in-plane polarization in Cartesian coordinates, are obtained by the transformations ${P_{\perp}(\bvec{s}) = P_{\perp}(g^{-1}\bvec{x})}$ and ${\D \bvec{P}_{\parallel}(\bvec{s}) = {g^{-1}}^T\D \bvec{P}_{\parallel}(g^{-1}\bvec{x})}$ and are shown in Figs.~\ref{fig:pol-main} (c) and (d), respectively. The shape of the out-of-plane polarization in bilayer hBN as a function of relative stacking is well known \cite{bennett2022electrically,bennett2022theory}: $P_{\perp}(\bvec{s})$ forms a triangular domain structure, with each domain having three neighboring domains of opposite polarization. The in-plane polarization has a remarkable structure: $\D \bvec{P}_{\parallel}(\bvec{s})$ flows into and out of the centers of the AB and BA domains, at which it is zero. The magnitude of $\D \bvec{P}_{\parallel}(\bvec{s})$ is maximal along the lines joining the centers of the AB and BA domains. A six-pointed star forms around the AA stacking configuration, at which both in-plane and out-of-plane components of the polarization are zero, modulo a quantum of polarization. Also, we note from symmetry that $\D \bvec{P}_{\parallel}(\bvec{s}) \propto \nabla_{\bvec{s}} P_{\perp}(\bvec{s})$.

\begin{figure*}[t!]
\centering
\includegraphics[width=\linewidth]{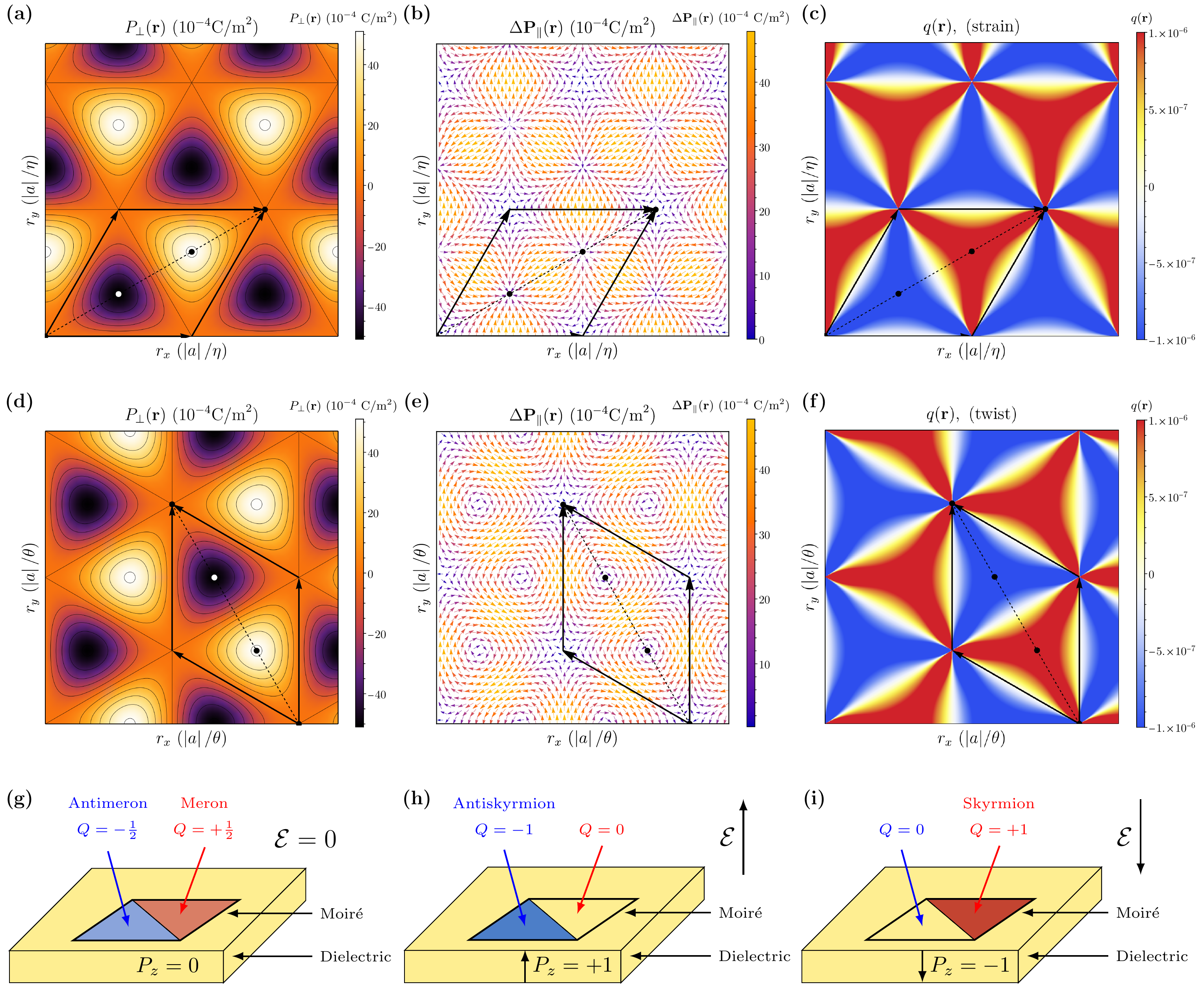}
\caption{\textbf{Polar topology in bilayer hBN.} \textbf{(a)}: out-of-plane polarization $P_{\perp}(\bvec{r})$, \textbf{(b)}: change in in-plane polarization $\bvec{P}_{\parallel}(\bvec{r})$ and \textbf{(c)}: local topological charge $q(\bvec{r})$ for biaxially strained hBN. \textbf{(d)}: out-of-plane polarization, \textbf{(e)}: change in in-plane polarization and \textbf{(f)}: local topological charge for twisted hBN. Plots are shown in real space $\bvec{r} = (r_x, r_y)$, with with length scales of the bilayer lattice constant $|a|$ divided by strain $\eta$ and twist angle $\th$ for strained and twisted bilayers, respectively. The black arrows represent the Moir\'e superlattice vectors in each case. \textbf{(g)-(i)}: Proposal of a system in which polar topological phase transitions may be driven by an applied electric field $\Ef$. A single cell of a moir\'e superlattice is embedded in a dielectric medium. \textbf{(g)}: at zero electric field, a meron-antimeron pair forms, i.e.~the topological charges in the individual domains are $Q=\pm \frac{1}{2}$. \textbf{(h)}: When a positive electric field is applied, the dielectric medium has normalized polarization $P_z=+1$, changing the winding around the boundary of the cell. As a result, the antimeron turns into an antiskyrmion ($Q\to -1$), and the meron vanishes ($Q\to 0$). \textbf{(i)}: When a negative field is applied, the reverse occurs: the meron turns into a skyrmion ($Q\to+1$), and the antimeron vanishes ($Q\to 0$).}
\label{fig:q-main}
\end{figure*}

Mapping from configuration space to real space, we see that the total polarization has a different form, depending on whether the moir\'e superlattice is induced via relative straining or twisting. For a relative strain $\eta$ between the layers, the mapping between configuration space and real space is $\bvec{s} = \eta\bvec{r}$. In this case, configuration space and real space are related via a simple scaling by $\eta$. In Fig.~\ref{fig:q-main} (a) and (b) we show the out-of-plane and in-plane polarization in a strained bilayer, which are identical to the polarization in configuration space. For a relative twist $\th$, the mapping between configuration space and real space is $\bvec{s} \approx \theta\begin{bmatrix*}[r] 0 & -1 \\ 1 & 0 \end{bmatrix*} \bvec{r}$, for $\th \ll 1$, (see Supplementary Material, Section I). For scalar fields, the quantities are related via a scaling by $\th$ and a reorientation of the cell vectors. For vector fields, the general form is different in both spaces. In Fig.~\ref{fig:q-main} (d) and (e) we show the out-of-plane and in-plane polarization in a twisted bilayer. We can see that the out-of-plane polarization has the same form, but the in-plane polarization curls in a clockwise manner around the centers of the AB/BA domains.

The components of polarization in Figs.~\ref{fig:pol-main} (a) and (b) were reproduced by integrating the dynamical charges, ${Z^{*}_{\k,\a\b} = V \frac{\partial P_{\a}}{\partial s_{\k,\b}} }$ \cite{ghosez1998dynamical}, (see Supplementary Material, Section V). This further allows the decomposition of the polarization into the contributions from the displacements of different atoms and in different directions. The polarization is mostly generated by in-plane sliding, with negligible contributions from the out-of-plane displacements, i.e.~the rippling of the interlayer separation as one layer slides over the other. As a result, the antisymmetric part of of $Z^{*}(\bvec{s})$ makes a significant contribution to the polarization, which is highly unusual for a ferroelectric material.

The in-plane component suggests that the polarization field in twisted bilayers does not just point in the out-of-plane direction, but exhibits intricate winding which is topologically nontrivial. Topology has played a manifest role in 2D materials, ranging from band theory to skymrions in magnetic systems. Such skymrions arise due to a mapping from a periodic unit cell to a classifying space that is topologically a sphere, as quantified via a homotopy $\pi_2(S^2)=\mathbb{Z}$ winding.

In Figs.~\ref{fig:q-main} (c) and (f) we show the local topological charge of strained and twisted bilayer hBN, respectively (see Methods). The AB and BA MPDs have equal and opposite winding. The total winding in each moir\'e cell is zero, but individually the AB and BA domains have winding numbers of $\pm \frac{1}{2}$, meaning the MPDs form a triangular network of merons and antimerons. The winding is concentrated at the domain centers and along the lines joining the AB/BA domain centers to the AA stacking configurations, and is zero along the domain walls. The magnitude of the winding is the same for strained and twisted bilayers apart from a reorientation of the axes, but the type of winding is different in each case. For strained bilayers the merons are of N{\'e}el type, where the polarization flows into and out of the domain centers. For twisted bilayers the merons are of Bloch type, where the polarization curls around the domain centers.

%The winding itself has an unusual form, with the most of the winding around the edges of the domains. In Fig.~\ref{fig:q-main} (b) we show the local winding in 1D along the $\hat{x}+\hat{y}$ unit cell diagonal. The winding is nonzero everywhere except along the domain walls. The winding at the domain centers, while nonzero, is roughly three orders of magnitude smaller than the winding near the AA stacking.

\section{Discussion}

In this work, we illustrate that, in addition to the out-of-plane polarization in layered systems like bilayer hBN, there is also an in-plane polarization as a result of a relative displacement between of the layers. This phenomenon is general to \textit{all} layered systems, provided that the bilayer lacks inversion symmetry. For the case of 3R-stacked hBN and similar materials (\chem{MoS_2}, etc.), there are three out-of-plane mirror planes related by $\Ct$ rotations plus an in-plane mirror plane half-way between the layers, various combinations of which are broken as one layer slides over the other.

Our findings indicate that the polar properties of layered systems are much richer than previously thought. For untwisted bilayer hBN, the energetically favorable AB and BA stackings have zero in-plane polarization. However, knowledge of how the in-plane polarization changes during the process of vdW sliding may prove useful. For example, by measuring the change in out-of-plane polarization, through a change in out-of-plane current, it is possible to determine when vdW sliding occurs between AB and BA domains, but it is not possible to determine in which direction the sliding occurs, i.e. to which of the three neighboring domains. By measuring the change in in-plane polarization, it may be possible to distinguish between these three sliding processes, which may enhance the capacity for information processing in ferroelectric layered systems.

Ferroelectric materials have been fabricated in many different geometries, from 2D thin films and FE/PE superlattices to 1D nanowires \cite{urban2002synthesis,yun2002ferroelectric} and nanotubes \cite{luo2003nanoshell,mao2003hydrothermal} (in fact, all nanotubes are inherently polar via flexoelectricity \cite{artyukhov2020flexoelectricity,springolo2021direct,bennett2021flexoelectric}), and 0D quantum dots \cite{chu2004impact,shin2005patterning}. Lower-dimensional ferroelectric systems typically exhibit size-dependent transitions in which the local polarization is more complex and can exhibit vortices before the polarization eventually vanishes completely \cite{fu2003ferroelectricity,naumov2004unusual,geneste2006finite,morozovska2006ferroelectricity}. Soon after, it was realized these polar structures with vortices were topologically nontrivial \cite{hong2010topology}, and skyrmion-like polarization structures were then identified, for example in barium titanate (\chem{BaTiO_3}) nanowires embedded in a matrix of strontium titanate (\chem{SrTiO_3}) \cite{nahas2015discovery}. It has also been proposed that skymrions may be created by controlling domains and domain walls in ferroelectrics, where at low temperatures the domain walls are of Bloch type and contain an in-plane polarization \cite{pereira2019theoretical}. Ferroelectric skyrmions have recently been experimentally observed in ferroelectric/paraelectric superlattices \cite{das2019observation,han2022high}. In addition to skyrmions, polar merons have also been considered theoretically and signaled experimentally \cite{wang2020polar}.

% different to oxide perovskites, dont form spontaneously, pinned by geometry, tuned by geometry

A full characterization of topological polarization also provides for an interesting avenue to explore. One major conceptual problem is that, unlike other topological invariants, the description of polarization in terms of exponentially localized Wannier functions requires topologically trivial electronic bands. It may be that topological polarization is a real space analogue to the topology of electronic bands in momentum space, with inversion symmetry and electric fields playing the role of time-reversal symmetry and magnetic fields. However, there is much work to be done in order to obtain a better understanding of topological polarization, and its relation to topological insulators in the presence of symmetries \cite{Qi_Rmp, Hasan_rmp2, Slager2013, bernevig2013topological}
as well as recently discovered multi-gap topological states due to the natural reality condition \cite{Bouhon2020, Unal_prl20202}. Nonetheless, it is evident that the topological polarization in moir\'e superlattices may serve as a new platform for topological physics in real materials, with great potential for the observation and control of topological phases in 2D materials.

We propose that it may be possible to drive a topological phase transition using a single moir\'e supercell embedded in a dielctric medium, from a single meron-antimeron pair into a skyrmion or antiskyrmion, for a fixed applied electric field, see Fig.~\ref{fig:q-main} (g)-(i). Such a setup may be possible by embedding a moir\'e quantum dot in a dielectric material, or in an aligned bilayer in which there is local strain or twisting around a defect, such that the bilayer is strained/twisted inside a domain and unstrained/untwisted outside. At zero electric field, a meron-antimeron pair forms in the supercell, and the polarization in the dielectric medium is zero. When an electric field is applied, the normalized polarization in the dielectric medium is $\pm \hat{z}$, which changes the winding along the boundary of the cell, adding to the winding in one domain, turning the meron/antimeron into a skyrmion/antiskyrmion, and canceling the winding in the other domain, making it topologically trivial, reminiscent of a bulk--boundary correspondence. A similar effect was also observed in periodic superlattices when the dielectric response of the system was taken into consideration. As a result, the polarization field has an in-plane component ${\epsilon_0\epsilon_{\perp}\Ef}.$ everywhere in the moir\'e superlattice, including the domain walls. This this leads to an induced winding along the domain walls, suggesting that even in periodic moir\'e systems, the merons/antimerons can be promoted to skyrmions/antiskyrmions, making the others topologically trivial, with an applied electric field.

It may also be possible to manipulate the meron--antimeron domain network via lattice reconstruction at different strains or twist angles. For small strains or twist angles, significant lattice relaxation can occur in order to increase the area of the more energetically stable stacking domains \cite{carr2018relaxation}, making the polar domains sharper \cite{bennett2022electrically,bennett2022theory}. At zero electric field, the AB and BA domains in 3R-stacked hBN relax evenly, leading to sharp triangular polar domains. However, the positions of the AA, AB, BA and SP stackings are all preserved. As a function of strain or twist angle, this is a continuous deformation which does not break any symmetries, and the meron--antimeron network should therefore be robust against lattice relaxation at zero electric field. When a field is applied, one type of domain will grow and the other will shrink, bending the polar domain walls. The robustness/fragility of the meron--antimeron network in response to the motion of the domain walls is not immediately clear. This goes beyond the scope of this work, however, and we leave it as a direction for future research.

In summary, we have illustrated that electronic out-of-plane and in-plane charge transfer and polarization are fundamental properties of layered systems without inversion symmetry. This will have far-reaching consequences both in terms of fundamental physics in strained/twisted and commensurate layered systems, such as ferroelectricity and topology, as well potential applications for ferroelectric-based nanodevices comprised of layers of 2D materials.

\section*{Methods} % Does not count toward word limit.
\subsection*{First-principles calculations}

First-principles density functional theory (DFT) calculations were performed using the {\sc siesta} \cite{siesta} and {\sc abinit} \cite{gonze2009abinit} codes, using {\sc psml} \cite{psml} norm-conserving pseudopotentials \cite{norm_conserving}, obtained from Pseudo-Dojo \cite{pseudodojo}. {\sc siesta} employs a basis set of numerical atomic orbitals (NAOs) \cite{siesta}, and double-$\zeta$ polarized (DZP) orbitals were used for all calculations. The basis sets in {\sc siesta} were optimized by hand, following the methodology in Ref.~\onlinecite{basis_water}. {\sc abinit} employs a plane wave basis set, which was determined using a kinetic energy cutoff of $1000$ eV. A mesh cutoff of $1200 \ \text{Ry}$ was used for the real space grid in all {\sc siesta} calculations. A Monkhorst-Pack $k$-point grid \cite{mp} of $12 \times 12 \times 1$ was used for the initial geometry relaxations, and a mesh of $18 \times 18 \times 1$ was used to calculate the polarization. Calculations were converged until the relative changes in the Hamiltonian and density matrix were both less than $10^{-6}$. In both codes, the revPBE exchange-correlation functional was used \cite{zhang1998comment}. The C09 \cite{dion2005erratum,cooper2010van} van der Waals correction was used in the {\sc siesta} calculations and the vdw-DFT-D3(BJ) \cite{becke2006simple} correction was used in {\sc abinit}. In {\sc siesta}, when an out-of-plane electric field was applied, a dipole correction \cite{dipole_correction_1,dipole_correction_2} was used in the vacuum region to prevent dipole-dipole interactions between periodic images. A dipole cutoff in slab-like systems has not been implemented in {\sc abinit}, so although a vacuum space of 50 \AA \ was used to separate the periodic images, the polarization is still slightly enhanced due to dipole-dipole interactions.

The top layer was translated along the unit cell diagonal over the bottom layer, which was held fixed. At each point a geometry relaxation was performed to obtain the equilibrium layer separation, while keeping the in-plane lattice vectors fixed. The out-of-plane and in-plane polarization were then obtained by calculating the Berry phases of the Bloch states. The data were fitted to Fourier expansions which respect the $\Ct$ rotation symmetry of bilayer hBN. It was found that both the out-of-plane and in-plane polarization were well described by the first order in the expansions, i.e. Eqs.~\eqref{eq:fit-Pz} and \eqref{eq:fit-Px}. At each point along the unit cell diagonal, DFPT calculations were performed using {\sc abinit} to calculate the dynamical charges.

\subsection*{Topological charge}

\begin{figure}[t!]
\centering
\includegraphics[width=0.75\linewidth]{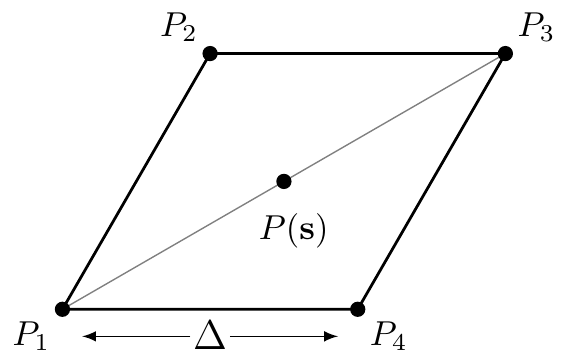}
\caption{\textbf{Winding with plaquettes.} Sketch of the plaquette defined around each $\bvec{P}(\bvec{s})$.}
\label{fig:plaquette}
\end{figure}

The winding number, or topological charge, of the polarization field in configuration space is
\beq{eq:Q-winding}
Q = \frac{1}{4\pi} \int \bvec{P}\cdot \lb \partial_{s_x} \bvec{P}\times \partial_{s_y} \bvec{P}\rb \dd\bvec{s}
\eec
which can mapped to real space in a strained or twisted bilayers as mentioned in the main text. Calculating the topological charge of a polarization field presents two additional complications when compared with magnetic fields. Firstly, the polarization is not of unit length, and must be normalized. Secondly, there are regions in space with zero polarization (modulo a quantum of polarization), i.e.~at the AA stacking configurations, near which Eq.~\eqref{eq:Q-winding} diverges. This can be avoided by calculating the topological charge following the methodology in Ref.~\cite{berg1981definition}.

The polarization in the unit cell is discretized on a fine grid with spacing $\D$. A plaquette is constructed around each grid point, see Fig.~\ref{fig:plaquette}. The plaquettes form a grid which is offset from the original by half a grid spacing (a similar technique is used in first-principles calculations for more efficient Brillouin zone integrations \cite{mp}). The zeros in polarization at the AA stacking configurations are thus not included in the offset grid. The local topological charge can then be defined as
\beq{}
q(\bvec{s}) = \frac{1}{4\pi}\lb A(P_1,P_2,P_3) + A(P_1,P_3,P_4) \rb 
\eec
where $A$ is the signed area spanned by three points on a sphere:
\beq{}
\begin{split}
A(P_1,P_2,P_3) = 2\arg & \big( 1 + P_1\cdot P_2 + P_2\cdot P_3 + P_3\cdot P_1 \big. \\
&+  \big. i P_1\cdot (P_2\times P_3) \big)
\end{split}
\eep
The total charge is then 
\beq{}
Q = \sum_{\bvec{s}} q(\bvec{s})
\eep
The total $Q$ in the configuration space unit cell sums to zero, with precision of around $10^{-12}$ even for relatively coarse grids. The winding numbers of the MPDs converge to $Q_{\text{AB}} = -Q_{\text{BA}} = \frac{1}{2}$ for grid spacings below $\D = 10^{-4}$, see Supplementary Fig.~5.

The meron--antimeron pair $\to$ (anti)skyrmion transition driven by an applied field was identified by calculating the topological charge using fixed boundary conditions outside the moir\'e cell rather than periodic boundary conditions. The normalized polarization in the cell is the same, and the normalized polarization outside the cell is taken to be $\text{sgn}\lb\Ef\rb$, where $\Ef$ is an applied field in the out-of-plane direction. The winding in the interior of the cell is unaffected, but an additional winding is induced along the boundary when a field is applied, see Supplementary Fig.~6. The total winding along the boundary sums to $-\text{sgn}(\Ef)$. For $\text{sgn}(\Ef) = \pm 1$, the boundary contributes an additional winding of $\mp \frac{1}{2}$ each to the meron/antimeron, promoting one to a skymrion/antiskyrmion, and making the other topologically trivial.

\section*{Acknowledgements}

D.~B.~acknowledges funding from the University of Li{\'e}ge under special funds for research (IPD-STEMA fellowship programme) and St.~John's College, University of Cambridge.
R.~J.~S and G.~C.~acknowledge funding from a New Investigator Award, EPSRC grant EP/W00187X/1. R.~J.~S also acknowledges funding from Trinity College, University of Cambridge.
E.~B.~acknowledges the FNRS and the Excellence of Science program (EOS ``ShapeME'', No. 40007525) funded by the FWO and F.R.S.-FNRS.
Ph.~G.~acknowledges financial support from F.R.S.-FNRS Belgium (grant PROMOSPAN) and the European Union’s Horizon 2020 research and innovation program under grant agreement number 964931 (TSAR).
The authors acknowledge the CECI supercomputer facilities funded by the F.R.S-FNRS (Grant No.~2.5020.1) and the Tier-1 supercomputer of the F\'ed\'eration Wallonie-Bruxelles funded by the Walloon Region (Grant No.~1117545). 

\section*{Author Contributions}

D.~B.~initiated the project, performed the first-principles calculations, calculated the topological charge and wrote the paper. E.~B.~and P.~G.~developed the understanding of polarization and ferroelectricity with D.~B., and provided guidance for the first-principles calculations. G.~C.~and R.~J.~S.~developed the understanding of topology with D.~B.~All authors discussed the results and contributed to the writing of the paper.

\section*{Data Availability}

The data presented in this study were generated using free and open source first-principles packages as described in the Methods section. The datasets generated during and/or analyzed during this study are available from the corresponding author upon request. 

\section*{Code Availability}

Code used to generate the plotted polarization structures is available from the corresponding author upon request.

\section*{Competing Interests}

The authors declare no competing interests.

%\bibliographystyle{naturemag}
%\bibliography{./references.bib}

\clearpage



\section*{Supplementary material}

Daniel Bennett,[1,2,3,*] Gaurav Chaudhary,[2] Robert-Jan Slager,[2] Eric Bousquet,[1] and Philippe Ghosez[1]

[1]*Physique Théorique des Matériaux, QMAT, CESAM, University of Liège, B-4000 Sart-Tilman, Belgium*
[2]*Theory of Condensed Matter Group, Cavendish Laboratory, University of Cambridge,*
*J. J. Thomson Avenue, Cambridge CB3 0HE, United Kingdom*
[3]*John A. Paulson School of Engineering and Applied Sciences,*
*Harvard University, Cambridge, Massachusetts 02138, USA*
(Dated: March 9, 2023)

## I. MEASURING POLARIZATION IN CONFIGURATION SPACE

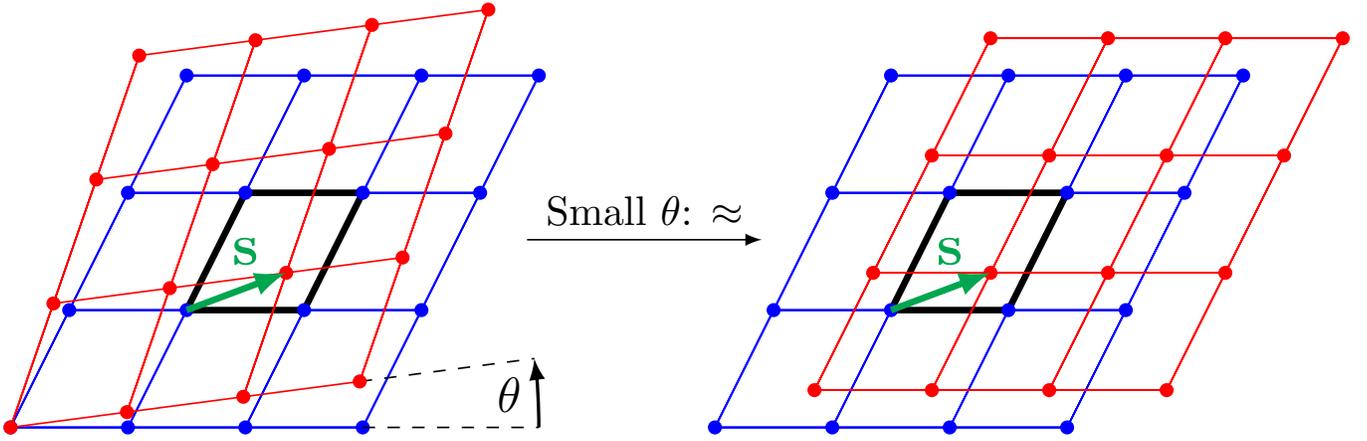

Supplementary FIG. 1. Sketch of how the configuration space mapping is used to estimate the local polarization in a twisted bilayer for small twist angles.

In this section we briefly describe how the configuration space mapping is used to estimate local properties in bilayers with small strains or twist angles. For a more detailed description, see Refs. 1 and 2. We illustrate the problem for the case of twist, considering two hexagonal monolayers (red and blue), each with a single atom in the unit cell, twisted at an angle $\theta$ with respect to one another, see Fig. 1. Suppose we want to estimate the local properties in the middle unit cell, marked in black, using first-principles calculations. Using the configuration space mapping, the relative position $\mathbf{s}$ of the red atom in this cell is $\mathbf{s}(\mathbf{r}) = (I - R_{\theta}^{-1})\mathbf{r}$, modulo any lattice vectors, where $\mathbf{r}$ is the real space position. For small twist angles, we make the approximation that the local changes in environment around the black unit cell are small, and the local properties can be described by a commensurate bilayer with a relative translation $\mathbf{s}$:

$$\mathbf{s} \approx \theta \begin{bmatrix} 0 & -1 \\ 1 & 0 \end{bmatrix} \mathbf{r} \, , \tag{1}$$

see Fig. 1. For a small homogeneous strain $\eta$, the equivalent mapping is

$$\mathbf{s} = \eta \mathbf{r} \, . \tag{2}$$

This allows the local properties in strained or twisted bilayers to be parameterized with first-principles calculations using a single commensurate cell of a bilayer, and sliding one layer over the other.

The total polarization, both out-of-plane and in-plane, can then be calculated from Berry phases:

$$\mathbf{P}(\mathbf{s}) = -\frac{1}{2\pi} \frac{e}{V(\mathbf{s})} \sum_{n}^{\text{occ}} \phi_{n,\alpha}(\mathbf{s}) \mathbf{a}_{\alpha} \, , \tag{3}$$

---

[*] dbennett@seas.harvard.edu

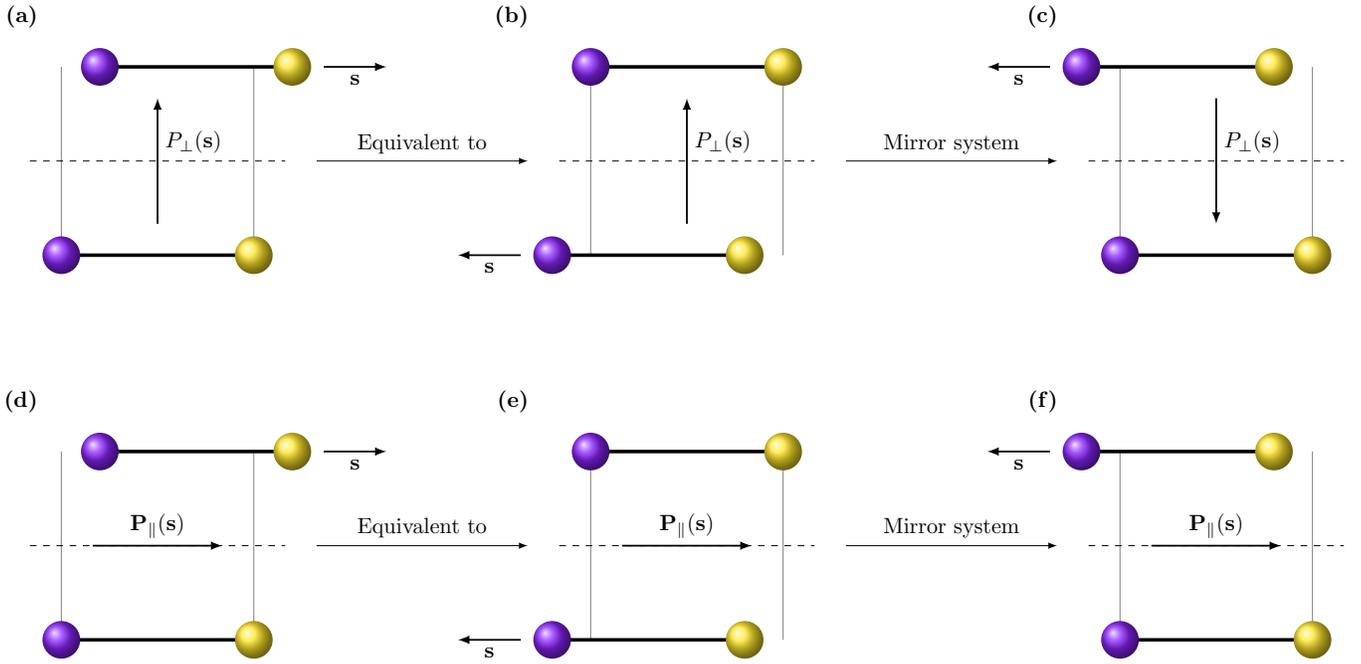

Supplementary FIG. 2. Sketch illustrating the effect of the mirror operation on **(a)-(c)**: the out-of-plane and **(d)-(f)**: the in-plane components of the polarization.

where $V(\mathbf{s}) = Ad(\mathbf{s})$ is the volume of the bilayer, $\mathbf{a}_\alpha$ are the in-plane lattice vectors, and $\phi_{n,\alpha}$ is the Berry phase. The sum is over all occupied bands $n$. The Berry phase is routinely calculated with most widely-available first-principles density functional theory (DFT) codes:

$$\phi_{n,\alpha}(\mathbf{s}) = \frac{(2\pi)^3}{V(\mathbf{s})} \int_{\text{BZ}} \langle u_{n,\mathbf{k}}(\mathbf{s}) | -i\mathbf{G}_\alpha \cdot \nabla_\mathbf{k} | u_{n,\mathbf{k}}(\mathbf{s}) \rangle \, \mathrm{d}\mathbf{k} \;, \tag{4}$$

where $\mathbf{G}_\alpha$ are the reciprocal lattice vectors and $u_{n,\mathbf{k}}(\mathbf{s})$ are the cell-periodic Bloch functions. Eq. (4) is essentially the position of the Wannier center, but normalized by the length of the in-plane lattice vectors and $2\pi$.

The local polarization in the main text was calculated by sliding one layer of hBN over the other, and at each point calculating the total polarization from Berry phases.

Of course, this method of calculating the local polarization in a twisted bilayer is not valid at all twist angles; at larger twist angles, the local changes in the environment cannot be assumed to be constant. We propose two more well-defined methods of measuring the local polarization:

1. By calculating the Wannier centers. The local polarization in each unit cell is then obtained by summing over the Wannier centers in that cell.

2. By integrating the dynamical charges.

In Section IV, we show that method 2 yields the exact same polarization obtained from Berry phases in configuration space. In principle, both of these methods should make it possible to directly calculate the local polarization in a twisted bilayer, although the calculations may be prohibitvely expensive.

## II. SYMMETRY OF THE POLARIZATION FIELD IN BILAYER hBN

Using the in-plane mirror symmetry of the AA stacking configuration, we can deduce the forms of the out-of-plane and in-plane polarization in 3R-stacked bilayer hBN.

For bilayer hBN with an AA stacking configuration, there is a mirror symmetry about the plane which is half-way between the two layers, and the out-of-plane polarization is zero. Suppose we translate the top layer by $\mathbf{s}$, generating an out-of-plane

polarization $P_\perp(\mathbf{s})$. This is equivalent to translating the bottom layer by $-\mathbf{s}$. If we now mirror the system about the plane between the two layers, the out-of-plane polarization is inverted. The same configuration can be achieved by sliding the top layer by $-\mathbf{s}$. Therefore we can deduce that

$$P_\perp(\mathbf{s}) = -P_\perp(-\mathbf{s}) \,, \tag{5}$$

i.e. the out-of-plane polarization is odd with respect to in-plane translations. This is illustrated in Figs. 2 (a)-(c).

Repeating this exercise with the in-plane polarization $\mathbf{P}_\parallel(\mathbf{s})$, we find that the in-plane polarization remains the same after applying the mirror symmetry, since it parallel to the mirror plane. Therefore we must have

$$\mathbf{P}_\parallel(\mathbf{s}) = \mathbf{P}_\parallel(-\mathbf{s}) \,, \tag{6}$$

i.e. the in-plane polarization is even with respect to in-plane translations. This is illustrated in Figs. 2 (d)-(f).

## III.   POLAR MODE PHONON FREQUENCIES IN BILAYER hBN

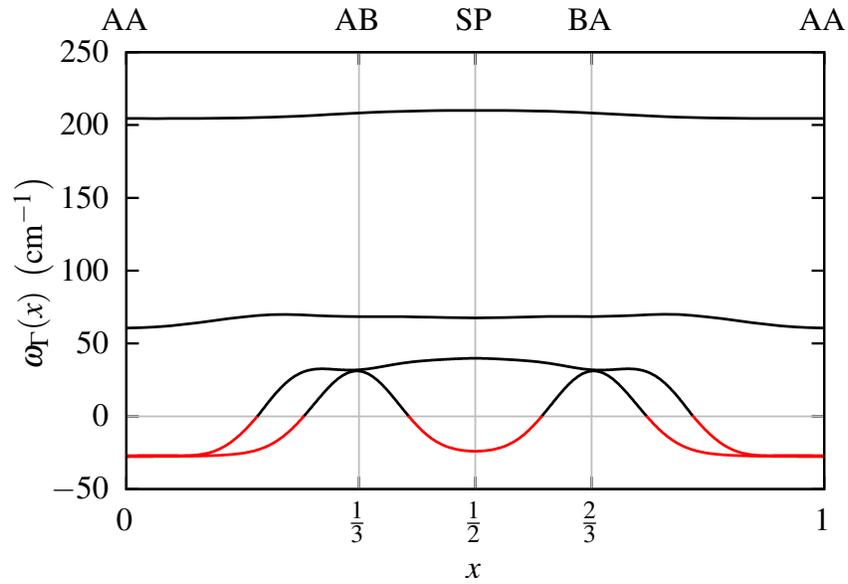

Supplementary FIG. 3. Lowest Gamma phonon frequencies in 3R-stacked bilayer hBN as a function of slide along the configuration space diagonal. The bands are shown in black when the phonons are stable, and red when they are unstable.

## IV.   DYNAMICAL CHARGES

In order to better understand the physical origin and nature of the in-plane and out-of-plane polarization, we calculated the dynamical charges[3] in bilayer hBN as a function of relative stacking:

$$Z^*_{\kappa,\alpha\beta} = V \frac{\partial P_\alpha}{\partial s_{\kappa,\beta}} \,, \tag{7}$$

i.e. the dipole generated in direction $\alpha$ in response to the unitary displacement of atom $\kappa$ in direction $\beta$. Since the bilayer is a semi-periodic system, Eq. (7) is measured at zero electric field for the in-plane responses, and with open circuit boundary conditions for the out-of-plane responses. Typically, Eq. (7) is symmetric or has a negligible anti-symmetric part, and can be diagonalized. The effective charge of each atom can therefore be thought of as an ellipsoid, with lengths and orientations of the axes determined by the eigenvalues and eigenvectors[3]. The change in polarization from a reference configuration at $\mathbf{s} = 0$ (AA stacking) to a general configuration at $\mathbf{s}$ can be measured by integrating the dynamical charges:

$$P_\alpha(\mathbf{s}) = \frac{1}{V(\mathbf{s})} \int_0^{\mathbf{s}} Z^*_{\kappa,\alpha\beta}(\mathbf{s}') \, \mathrm{d}s'_{\kappa,\beta} \,. \tag{8}$$

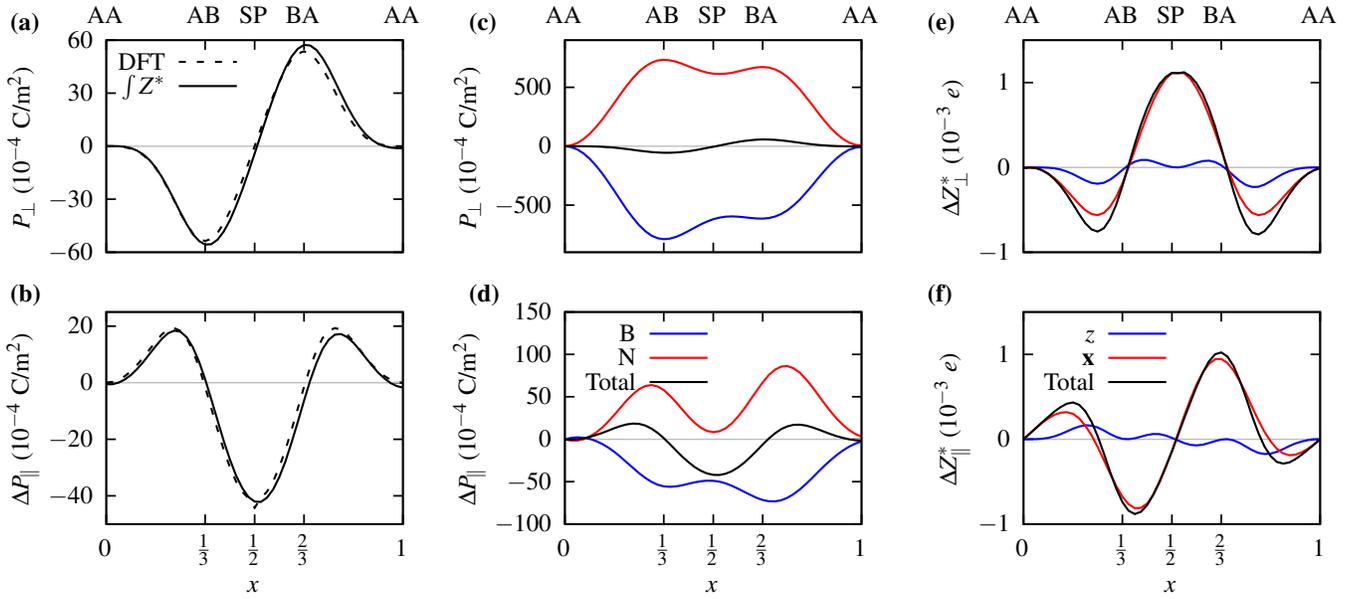

Supplementary FIG. 4. Total polarization in **(a)** the out-of-plane direction and **(b)** the in-plane direction, obtained by integrating the dynamical charges (solid), and from Berry phases (dashed). Individual contributions to the total polarization (black) in **(c)** the out-of-plane direction and **(d)** the in-plane direction, from the displacement of the B (blue) and N (red) atoms. Total change in dynamical charges (black) and the decomposition into out-of-plane (blue) and in-plane (red) displacements in **(e)** the out-of-plane direction and **(f)** the in-plane direction.

The dynamical charges were calculated for bilayer hBN along the diagonal in configuration space using density functional perturbation theory (DFPT) calculations in ABINIT. Integrating the effective charges results in a polarization almost in exact agreement with the polarization obtained by calculating the Berry phases, see Figs. 4 (a) and (b).

One advantage to calculating the polarization from the effective charges is that it allows the decomposition of the total polarization into contributions from the displacement of individual atoms in different directions. In Figs. 4 (c) and (d), the separate contributions to the polarization arising from the displacement of the B and N atoms in the top layer are shown. For the out-of-plane polarization, displacing the two atoms generates large, almost equal but opposite polarizations, which cancel to give the dipole generated by the transferred electronic charge. The individual contributions to the in-plane polarization have the same order of magnitude as the total in-plane polarization. Neither contribution by itself is even, but their sum is even. The total change of the effective charges in the top layer are shown in Figs. 4 (e) and (f), from which it is clear that the sum of the effective charges is the integrand of the total polarization. Notably, the polarization is almost entirely generated from the relative in-plane displacements. Out-of-plane displacements, i.e. the modulation of the interlayer separation, results in a higher-order contribution which is much smaller in magnitude.

## V.  CONVERGENCE OF THE TOPOLOGICAL CHARGE

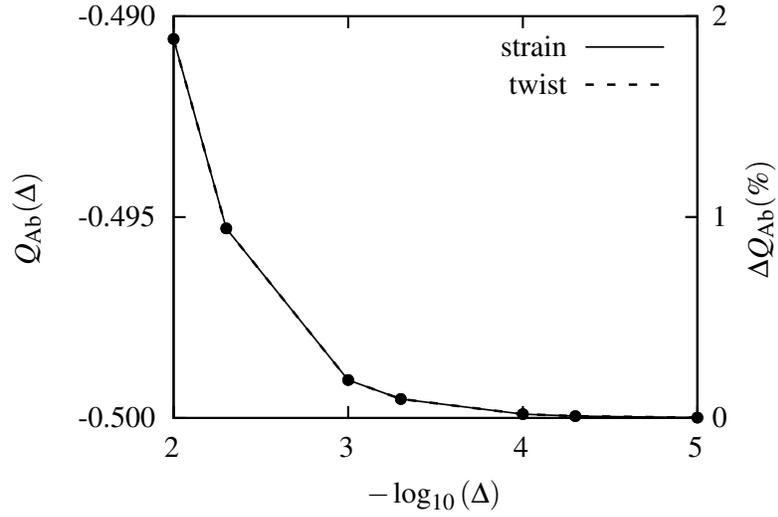

Supplementary FIG. 5.  Convergence of the total topological charge / winding number of the AB domain $Q_{\mathrm{AB}}$ for a strained (solid line) and twisted (dashed line) bilayer as a function of grid spacing $\Delta$. The second $y$-axis shows the percentage error.

## VI.  WINDING ALONG THE CELL BOUNDARIES FOR AN APPLIED ELECTRIC FIELD

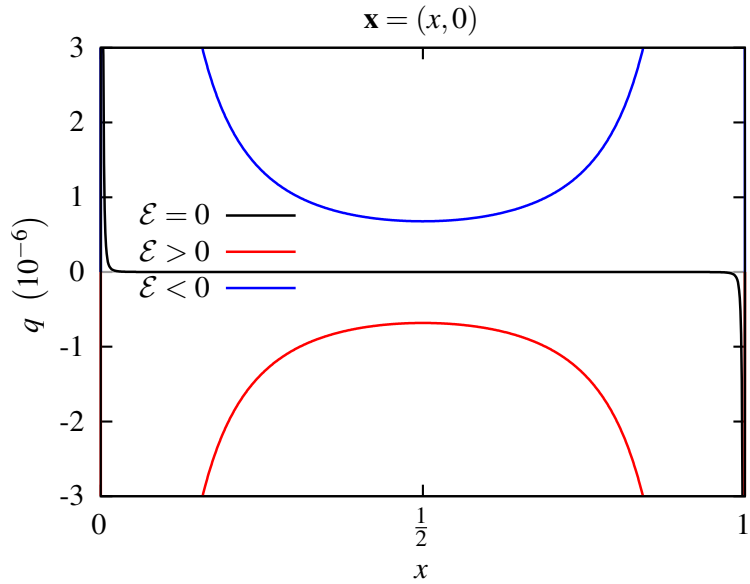

Supplementary FIG. 6.  Local winding along $\mathbf{x} = (x, 0)$ in a single moiré cell embedded in a dielectric medium for zero field (black) and for out-of-plane electric fields applied in the positive (red) and negative (blue) directions.

**SUPPLEMENTARY REFERENCES**

---



\end{document}